\documentstyle[12pt]{article}

\begin{document}

\def\lp{\lambda^\prime}
\def\lpp{\lambda^{\prime \prime}}
\def\lbf{\lambda^{\prime \prime}_{133}}
\def\lbs{\lambda^{\prime \prime}_{233}}
\def\lbv{\lambda^{\prime \prime}_{B\!\!\!/}}
\def\lbi{\lambda^{\prime \prime}_{i33}}
\def\abv{A^0_{B\!\!\!/}}
\newcommand{\be}{\begin{equation}}
\newcommand{\ee}{\end{equation}}
\newcommand{\bea}{\begin{eqnarray}}
\newcommand{\eea}{\end{eqnarray}}
\begin{center}
\large {\bf A See-saw model of sterile neutrino}
\\ 
\vskip 1in Biswajoy Brahmachari \\
\end{center}
\begin{center}
Department of Physics\\
Indiana University, Bloomington\\
IN-47405, USA
\\
\end{center}
\vskip 1in
{
\begin{center}
\underbar{Abstract} \\
\end{center}
{\small 
If the smallness of the mass of the sterile neutrino is to be explained 
by the see-saw mechanism, the off-diagonal entries of the mass matrix
needs to be protected by some symmetry not far above the electroweak
scale. We implement see-saw mechanism in a gauge model based on $SU(2)^q_L
\times SU(2)^l_L \times U(1)^q_Y \times U(1)^l_Y$ un-unified gauge group
which breaks to $SU(2)_L \times U(1)_Y$ at the TeV region via a two-step
symmetry breaking chain. The right handed diagonal block is tied to the
highest scale up to which the un-unification symmetry holds. The sterile
neutrino emerges from a quark-lepton mixed representation of the
un-unified group.} 

\newpage

\noindent 
A light sterile neutrino ($\nu_s$) is necessary to explain the 
solar\cite{solar}, atmospheric\cite{atmospheric} and LSND\cite{lsnd} 
neutrino anomalies simulteniously via neutrino mixing schemes. The 
sterile neutrino has to be approximately degenerate in mass with either 
$\nu_e$ or $\nu_\mu$ and it must have a mixing angle compatible with those 
found by the oscillation experiments. Neutrinos in the eV mass range can
also play the role of hot dark matter\cite{hdm} of the universe. If solar 
neutrino anomaly is explained through the matter-induced oscillation $\nu_e 
\leftrightarrow \nu_s$ the corresponding mixing angle $\sin^2 2
\Theta_{es}$ can either be $10^{-2}$ or of the order of unity 
with $\Delta~m^2_{es} \sim 10^{-5}$ eV$^2$. In this case
the atmospheric neutrino anomaly has to be explained through
maximal $\nu_\mu \leftrightarrow \nu_\tau$ mixing. On the contrary if
solar neutrino anomaly is explained through $\nu_e \leftrightarrow
\nu_\tau$ oscillation, the atmospheric neutrino anomaly has to be
explained through maximal $\nu_\mu \leftrightarrow \nu_s$ oscillation  
with $\Delta~m^2_{\mu s} \sim 10^{-3}$ eV$^2$. The invisible decay width
of the $Z$ boson work oneself into a position that there are three
neutrinos\cite{neu} coupling to the $SU(2)_L \times U(1)$ invariant weak
neutral current. Fourth neutrino has to be a singlet under the standard
model gauge symmetry, or in other words, it has to be sterile.

A natural way to get a small neutrino mass is via the see-saw 
mechanism\cite{seesaw}. In this case the mass matrix can formally 
be written as,
\be
\bordermatrix{& \nu_L & \nu_R \cr
   \nu_L & 0 & m_D \cr
    \nu_R & m_D & M_X\cr
}.
\ee
Note that the off-diagonal Dirac type mass is $G \equiv SU(2)_L \times
U(1)_Y$ symmetry breaking. This makes the off diagonal entry  
of the order of $m_Z$. The diagonal entry, however, is $G$ conserving 
and can be taken to be a large scale of the order of the right-handed 
symmetry breaking scale ($M_R$) or the GUT scale ($M_X$). We have
generically termed this scale $M_X$ later. The matrix has two eigenvalues 
$m^2_D/M_X$ and $M_X$. The first eigenvalue explains the smallness of the 
neutrino mass when $M_X \rightarrow \infty$. If on the 
other hand the off-diagonal entry is also $G$ conserving, the mass
eigenvalues will be of the order of $M^2_X/M_X = M_X$ and $M_X$. Obviously
in this case see-saw mechanism cannot explain the smallness of neutrino
mass. This situation arises in the case of a singlet or sterile neutrino
as in this case  the off-diagonal elements originate from singlet Higgs
scalars. In this paper our purpose is to study a minimally modified
version of the standard model gauge symmetry $G$ and construct a model for
the sterile neutrino in such a way that a naturally light sterile neutrino
as well as the required mixing angles with the ordinary species can be
explained through see-saw mechanism.

As we stick to the see-saw mechanism, the Dirac type off-diagonal entry of
the sterile neutrino have to be protected near the electroweak scale by
some symmetry\cite{bm}. In this paper we will consider a variation on an 
ingenious gauge model based on quark-lepton un-unification 
symmetry \cite{gjs,ma} 
\be
G^0=SU(q)^q_L \times SU(2)^l_L \times U(1)_Y,
\ee
which breaks to 
standard model gauge group at a scale $M_E \sim O(TeV)$. Our gauge group
was studied in Ref(\cite{dbchou}) and we will see that it will nicely 
fits to our purpose. The
breaking chain is, 
\begin{eqnarray}
&&G^\prime~=~SU(2)^q_L \times SU(2)^l_L \times U(1)^q_{Y} \times 
U(1)^l_Y, \nonumber\\ 
M_X \longrightarrow && G^0~=~SU(2)^q_L \times SU(2)^l_L \times U(1)_{Y} 
\nonumber\\
M_E \longrightarrow && G~=~SU(2)_L \times U(1)_{Y}. 
\end{eqnarray}
When the electric charge is expressed in terms of the generators 
of $G^{\prime}$, we get
\begin{equation}
Q= T^3_q + T^3_l + Y^q + Y^l.
\end{equation}
The global fits of all electroweak precision parameters put limit on the 
mass scales of the extra gauge bosons belonging to group $G^0$. The
model based on $G^0$ has been studied in literature extensively. In 
Ref\cite{chiv} it has been shown that the additional gauge bosons 
should be heavier than 2 TeV depending on the new mixing angle of the
un-unified group defined similar to the weak mixing angle of the standard 
model. The heavier gauge bosons will induce additional box diagrams
contributing to $B_0-\overline{B_0}$ mixing. Furthermore the deviation of
$e^+~e^-\rightarrow \mu^+~\mu^-$ and $e^+~e^- \rightarrow b~\overline{b}$
asymmetries from the standard model predictions restrict the mixing of
ordinary and extra gauge bosons \cite{rizzo2, randall, rizzo}. 
In this letter, we will simply choose the VEV of $H_E=2$ TeV as a
tentative value inspired by \cite{chiv} varying the Yukawa couplings. We
could have taken a different value of $<H_E>$ which will yeald a separate
set of Yukawa couplings. Quark mass generation in this model is
complicated. An approach is sketched in Ref\cite{gjs}. 

We add extra fermions $S_L=(2,2,1/2, -1/2)$ and their right handed singlet 
counterparts $S_R$. $S_L$ contains a singlet of $G$ which will be the
left-handed partner of our sterile neutrino. The right handed
sterile neutrino is a singlet of $G^\prime$\footnote{One can
consider an even more baroque but symmetric scheme such as SU(16)
\cite{su16} where the right handed sterile neutrino also emerges from the 
bi-doublet representation}. Furthermore a Higgs
scalar field (2,2,1/2,-1/2) is needed to break the group $G^\prime$. This
is the lowest dimensional non-trivial representation which contains a
singlet of $G$. 

The model based on group $G^\prime$ is theoretically incomplete. The 
standard model particle content introduces triangle anomalies as separate 
anomalies related to the quark and leptonic parts do not add up to 
zero individually. A prescription of additional fermions in similar 
models is forwarded in References\cite{gjs,rizzo}. We will give the 
details of extra fermions which cancell anomaly and their effects on 
quark masses and flavor changing neutral currents in along the lines of
\cite{dbchou} in a future publication. Let us summarize the particle 
content. The fermions and scalars transform in the notation $G^\prime 
\rightarrow G$ as, 
\begin{eqnarray} 
&&FERMIONS \nonumber\\ &&\nonumber\\
Q_L &=& (2,1,1/6,0) \rightarrow (2,1/6)\nonumber\\
U_R &=& (1,1, 2/3,0) \rightarrow (1,2/3) \nonumber \\
D_R &=& (1,1,-1/3,0) \rightarrow (1,-1/3) \nonumber\\
L_L &=& (1,2,0,-1/2) \rightarrow (2,-1/2) \nonumber\\
E_R &=& (1,1,0,1) \rightarrow (1,1) \nonumber \\
&& \nonumber\\
&& EXTRA~FERMIONS \nonumber\\
&& \nonumber\\
N_R &=& (1,1,0,0) \rightarrow (1,0) \nonumber \\
{S_L }&=& {(2,2,1/2,-1/2)} \rightarrow (3,0) + (1,0)\nonumber\\
{S_R } &=& {(1,1,0,0)} \rightarrow (1,0) \nonumber\\
&& \nonumber \\
&& HIGGSES \nonumber\\
&& \nonumber\\
H &=& (1,2,0,1/2) \rightarrow (2,1/2)\nonumber\\
H_E&=&(2,2,0,0) \rightarrow <H_E> \sim M_E \nonumber\\
H_X&=&(1,1,1/2,-1/2) \rightarrow <H_X> \sim M_X \nonumber
\end{eqnarray}
Note that the bi-doublet has a triplet and a singlet at low energy.
The singlet is interpreted as the left handed sterile neutrino. Given the
particle
spectrum it is easy to construct the neutrino mass matrix in terms of the
VEVs of the fields,
\be
<H_X>=\eta~~,~~<H_E>=\sigma~~,~~<H>=v. 
\ee
For this toy model it is enough to consider only the electron 
generation. We will check whether a $\nu_e \leftrightarrow \nu_s$ 
solution of the solar neutrino problem is possible. The Atmospheric and 
LSND solutions will depend on the Yukawa matrix of the three active 
neutrino species\footnote{We have checked that it is also possible to get 
the $\nu_s \leftrightarrow \nu_\mu$ solution to the atmospheric neutrino 
anomaly in a similar manner.}. The simplified 
neutrino mass matrix is 
\be \bordermatrix{& \nu_L & S_L & N_R & S_R \cr
   \nu_L & 0 & 0 & h_1 v & h_2 v \cr
    S_L & 0 & 0 & h_3 \sigma {\eta \over M_X} & h_4 \sigma {\eta \over
     M_X} \cr
    N_R & h_1 v & h_3 \sigma {\eta \over M_X} & O(M_X) & O(M_X) \cr
    S_R & h_2 v & h_4 \sigma {\eta \over M_X} & O(M_X) & O(M_X) \cr
} \label{mat}
\ee
The non-renormalizable Yukawa couplings $h_3$ and $h_4$ are
interesting. They keep the effective strength of the off-diagonal
elements at the TeV region. The light neutrino masses are the two light
eigenvalues of Eqn.(\ref{mat}). It is easy to see that they are also the 
eigenvalues of the light neutrino mass matrix  given by 
\be 
m_{light}=m_{Dirac}~{1 \over M}~m^\dagger_{Dirac} 
\ee
where, symbolically we have expressed $m_{dirac}$ as the off-diagonal 
$2\times 2$ block and $M$ as bottom-right $2 \times 2$ block of the 
matrix in Eqn.(\ref{mat}). We will further assume
\be
M=M_X \pmatrix{ h_5 & h_6 \cr h_7 & h_8}
\ee
Now we are in a position to give the results. We take parameters as,
\be
v=256 {\rm GeV},~~\sigma= 2000 {\rm GeV},~~M_X=10^{16}~~{\rm GeV}
\ee
and then try to find natural values of Yukawa couplings in the 
range $ 10^{-3} - 1$ which gives $\Delta~m^2 \sim 10^{-5}$ and 
$\sin^2 ~2 \Theta \sim 1-2\%$. Note that $<\sigma> = 2$ TeV is a choice 
value and should only be looked upon a sample value in the TeV range. An
exact value of $<\sigma>$ is not required for our purposes. A set of
solution in given in Table (\ref{table1}). It's not the precise values in
table (\ref{table1})  that are the "results", but the concept they
represent (one can avoid fine-tuning while providing sterile neutrinos).  
\begin{table}[htb]
\begin{center}
\[
\begin{array}{|cccccccc||c||c|}
\hline
h_1 & h_2 & h_3 & h_4 & h_5 & h_6 & h_7 & h_8 
& \sin^2~2
\Theta  & \Delta~m^2              \\ 
\hline
0.1 & 0.05 & 0.05 & 0.1 & 1.0 & 0.4 & 0.4 & 1.0 & 0.022 & 10^{-5.1} \\   
0.1 & 0.04 & 0.04 & 0.1 & 1.0 & 0.4 & 0.4 & 1.0 & 0.010 & 10^{-5.1} \\
0.1 & 0.04 & 0.04 & 0.1 & 1.0 & 0.3 & 0.3 & 1.0 & 0.016 & 10^{-5.2} \\
\hline
\end{array}
\]
\end{center}
\caption{A set of natural values of neutrino Yukawa couplings giving
desired 
masses and mixing} 
\label{table1}
\end{table}

We note that in the standard model the $U(1)_Y$ symmetry protects the 
lepton masses from shooting up to the Plank scale as all representations 
of the SU(2) group are self-conjugate. Had we considered the group
$G^0=SU(2)^q_L \times SU(2)^l_L \times U(1)_Y$ istead of $G^\prime$ and 
introduced the extra fermion $S_L=(2,2,0)$ under $G^0$ we would have had 
the same consequence. The $S_L~S_L$ entry of the mass matrix in Eqn 
(\ref{mat}) wouldn't have been protected around the TeV scale. This is 
the justification of using $G^\prime$ in this paper.

To conclude, the standard model is contructed in such a way that the
neutrino remains massless. If we want to have a neutrino mass in the eV
range from the VEV of the standard model Higgs doublet we need to add a
right handed neutrino and the corresponding `Dirac type' neutrino Yukawa 
coupling needs to be fine tuned to the precision of
$10^{-9}/10^2=10^{-11}$. A natural solution to this is the see-saw
mechanism where the mass of the light neutrino emerges as
$m^2_{weak}/M_{X}$ from the diagonalization of a `Majorana-type' mass
matrix. We obtain a light neutrino in the eV range when $M_X \sim
10^{13.8}$ GeV, where $M_R$ is some large scale of the theory. However the
sterile neutrino, being a  singlet, does not feel the effect electroweak
symmetry breaking as it does not couple to the Higgs doublet which breaks 
the electroweak symmetry. All mass scales relevant to the
sterile neutrino shoots off to the largest scale $M_X$ up to which the
standard model symmetry remains exact. Thus the see-saw mechanism breaks
down. This is because see-saw mechanism necessarily needs an interplay
between two scales, the weak scale and the scale $M_X$ in the present
circumstances.

In this note we have constructed a scenario where the quark lepton 
un-unification symmetry exists near the TeV scale and this symmetry 
gives the required protection to the Dirac type off-diagonal mass of a
sterile neutrino, which is embedded in a bi-doublet representation of
the un-unified group. Hence in this scenario the sterile neutrino is a low 
energy manifestation of a quark-lepton `bi-doublet' mixed representation
which feels the effect of the breaking of the un-unified symmetry. 
We assume that the VEV $\sigma$ which breaks the un-unification group is
at the TeV range whereas the quark-lepton U(1) groups (U(1)$^l$  and
U(1)$^q$) merge at the high scale $M_X$. We have used $natural~values$ of
all the Yukawa couplings ($0.001 - 1.0$) and succeeded in obtaining the
feasible mass scale of the sterile neutrino and its mixing with the
electron neutrino in the context of the solar neutrino anomaly. Thus, in
doing so we did not need to fine-tune the Yukawa couplings. This may
explain the solar neutrino anomaly. In doing this we have used a variation
of the see-saw mechanism.

\vskip 1cm

It is a pleasure to acknowledge discussions with M. S. Berger and 
G. Senjanovi\'c.

\end{document}